\documentclass[conference]{IEEEtran}
 
\usepackage{amsthm,amsmath,amssymb}

\usepackage{mathrsfs}
\usepackage{amsfonts}
\usepackage{graphicx, amssymb}
\usepackage{amsthm}
\usepackage{amsmath}
\usepackage{epsfig}
\usepackage{graphicx}
\usepackage{graphics}
\usepackage{multirow}
\usepackage{threeparttable}
\usepackage{amsthm}
\usepackage{float}
\usepackage{color}
\usepackage{algorithm}
\usepackage{algpseudocode}
\usepackage{bm}
\usepackage{algorithmicx}  
\usepackage{algpseudocode}
\usepackage{setspace}
\usepackage{authblk} 
\usepackage{cite}

\usepackage{xcolor}

\newcommand{\RNum}[1]{\uppercase\expandafter{\romannumeral #1\relax}}

\ifCLASSINFOpdf
\else
\fi

\begin{document}

\title{Cell-Free Symbiotic Radio:  Channel Estimation Method and Achievable Rate Analysis }

%

\author{
   {Zhuoyin Dai{*}, Ruoguang Li{*}, Jingran Xu{*}, Yong Zeng{*$\dagger$}, and Shi Jin{*} }  \\
  {{*}National Mobile Communications Research Laboratory, Southeast University, Nanjing 210096, China} \\
  {{†}Purple Mountain Laboratories, Nanjing 211111, China} \\
  {Email: \{zhuoyin\_dai, ruoguangli, jingran\_xu, yong\_zeng, jinshi\}@seu.edu.cn}
}
\maketitle


\begin{abstract}
Cell-free massive MIMO and symbiotic radio are promising beyond 5G (B5G) networking 
architecture and transmission technology, respectively.  This paper studies 
cell-free symbiotic radio systems, where a number of distributed access points (APs) 
cooperatively send primary information to a receiver, and simultaneously support 
the backscattering communication of the secondary backscatter device (BD). 
An efficient two-phase uplink-training based channel estimation method is proposed
to estimate the direct-link channel and cascaded backscatter channel,
and the achievable primary and secondary communication rates taking
into account the channel estimation errors are derived.  
Furthermore, to achieve a flexible trade-off between the primary and secondary 
communication rates,  we propose a low-complexity \emph{ weighted-maximal-ratio transmission} 
(weighted-MRT)  beamforming scheme, which only requires local processing at 
each AP without having to exchange the estimated channel state information. 
Simulation results are provided to show 
the impact of the channel training lengths on the performance of 
the cell-free symbiotic radio systems.

\end{abstract}


\IEEEpeerreviewmaketitle

\vspace{0ex}
\section{Introduction}
\vspace{0ex}
\IEEEPARstart{A}{long} with the rapid deployment of the fifth-generation (5G) 
mobile communication networks, researchers have started the investigation of 
6G targeting for network 2030 \cite{LavtaM2019a,YouX2021a}. In order to support  
orders-of-magnitude performance improvement  in terms of coverage, 
connectivity density, data rate, reliability, latency, etc., many 
promising technologies have been extensively studied, such as
extremely large-scale MIMO/surface \cite{HuS2018a,LuH2021a}, TeraHertz communication \cite{ElayanH2018a }, 
non-terrestrial networks (NTN) \cite{GiordaniM2021a, YZeng2019a }, 
and AI-aided wireless communications \cite{LuongNC2019a}.
In particular, 
cell-free massive MIMO \cite{Ngo2017a} and symbiotic radio \cite{YLiang2020a} 
were recently proposed as promising networking 
architecture and transmission technology for beyond 5G (B5G), respectively.

Cell-free massive MIMO is different from the classical cellular networking 
architecture in the sense that it blurs  the conventional concepts of 
cells or cell boundary \cite{Ngo2017a}. Instead, distributed 
access points (APs), which are connected to the central processing unit (CPU), 
exploit their local channel state information (CSI) to simultaneously serve the 
users. 
As  such, cell-free massive MIMO system is expected to mitigate the inter-cell
interference issues in small cell 
systems and  provides users with appealing uniform good service 
everywhere \cite{Interdonato2019a}. 
Meanwhile,   no exchange of CSI is required among different APs, which enables 
low complexity and light backhaul load between APs and CPU.  
Therefore, significant research efforts have been recently devoted to
the theoretical analysis and practical design of cell-free massive MIMO, 
e.g., precoding design 
\cite{Nayebi2017a}, power optimization \cite{Ngo2017a}, and energy 
efficiency analysis \cite{NguyenL2017a,Ngo2018a}.

On the other hand, in terms of transmission technology for B5G, 
symbiotic radio,   which combines the benefits of the 
conventional cognitive radio (CR) and ambient backscattering 
communications (AmBC),  has been proposed for spectral- and 
energy-efficient communications \cite{YLiang2020a}. 
In typical symbiotic radio systems, the secondary device not 
only utilizes the spectral but also the power of the primary 
system via the passive backscattering technology \cite{RLong2019a}. 
Based on the relationship of symbol durations of the primary
and the secondary signals,  symbiotic radio system can be classified as
\emph{commensal symbiotic radio} (CSR) and \emph{parasite symbiotic radio} (PSR) \cite{RLong2020a}. 
In CSR, the secondary signals have  much longer symbol duration 
than the primary signals, making the secondary backscattering transmission 
contribute additional multipath components to enhance the primary communication. 
As a result, the primary and secondary communications form a mutualism 
relationship \cite{RLong2020a}.  By contrast, in PSR,
the primary and secondary signals have equal symbol duration,
and the secondary signals are often treated as interference to the primary signals.
Significant research efforts have been recently devoted to the study of symbiotic radio systems, 
e.g., in terms of performance analysis \cite{WuT2021a} and resource allocations \cite{GuoH2019a, ChuZ2020a}.


However, all the aforementioned existing works studied cell-free massive MIMO and 
symbiotic radio separately, i.e., cell-free  system with the conventional 
active transmission or symbiotic radio transmission in conventional 
cellular network or  the basic point-to-point communications. As the promising 
B5G networking architecture and transmission technology, respectively, it is 
natural that cell-free networking and symbiotic radio communication would 
merge each other to reap the benefits of both. This thus motivates our current 
work to study cell-free symbiotic radio systems, which, to our best knowledge, 
have not been studied in the existing literature.   
In cell-free symbiotic radio systems, a number of distributed  APs
cooperatively send primary information to a receiver, and concurrently support 
the passive backscattering communication of the secondary backscatter device (BD). 
As such,  the distributed cooperation gain by APs can be exploited to 
enhance both the primary and secondary communication rate.  
An efficient two-phase uplink-training based channel estimation method 
is proposed to estimate the direct-link channel and cascaded backscatter channel, 
respectively.  
Furthermore, the interrelationship between the primary and secondary transmission
is revealed by deriving their achievable rates taking into account the  
channel estimation errors.  Besides, to achieve a flexible trade-off 
between the primary and secondary 
communication rate, a low-complexity {\it weighted-maximal-ratio
transmission} (weighted-MRT) beamforming scheme is proposed, which only requires 
local processing at each AP without having to exchange the estimated CSI among APs. 
Numerical results are provided to show the performance of the cell-free symbiotic 
radio system with different training lengths.


\section{System Model}
As shown in Fig. 1, we consider a cell-free symbiotic radio system, which 
consists of  $M$ distributed  APs, one information receiver, and one BD. 
The $M$ APs   cooperatively send primary information to the receiver,  
and simultaneously  support the  BD for secondary communication via 
backscattering to the same receiver.  The considered setup may model
a wide range of  applications, e.g., with the 
receiver corresponding to smartphones and the BD being the smart 
home sensor nodes.  
\begin{figure}
  \setlength{\abovecaptionskip}{-0.1cm}
  \setlength{\belowcaptionskip}{-0.3cm}
	\centering
\includegraphics[height=1.9in, width=2.46in]{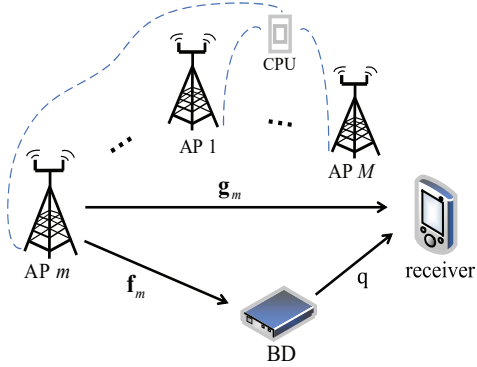}
\caption{Cell-free symbiotic radio, where $M$ distributed  APs cooperatively 
transmit primary information to the receiver and concurrently  support the
 secondary backscattering communication.    }
  \vspace{-0.4cm}
\end{figure}
We assume that each AP is equipped with $N$ antennas, and the receiver and BD each has 
one antenna.  Denote by $\mathbf{g}_m\in \mathbb{C}^{N\times 1}$ and 
$\mathbf {f}_m\in \mathbb{C}^{N\times 1}$
the  multiple-input single-output (MISO) channels from the $m$th AP to the receiver and BD, 
respectively, where $m=1,...,M$. Further denote by $q\in \mathbb{C}$  the channel coefficient 
from the BD to the receiver. Then, the  cascaded backscatter channel
from the $m$th AP to the receiver via the BD is $q\mathbf{f}_{m}$.

In this paper, we focus on the PSR setup\cite{RLong2020a},
where the symbol duration of the primary and secondary signals are equal. Let
$s(n)\sim\mathcal{CN}(0,1)$ 
and $c(n)\sim\mathcal{CN}(0,1)$   denote the circularly symmetric complex Gaussian (CSCG) 
information-bearing symbols of the primary 
and secondary signals respectively. 
Further denote by $p$ the transmit power of each AP, and $\mathbf {w}_{m} \in 
\mathbb{C}^{N\times 1}$ with $\|\mathbf {w}_{m}\|^{2}= 1$ denotes the transmit beamforming vector 
of the $m$th AP.
Then the received signal at the receiver  is 
  \begin{equation}
\setlength\abovedisplayskip{2pt}
\setlength\belowdisplayskip{2pt}
    r(n)= \!\sum_{m=1}^{M} \!\big[ \sqrt{p}\mathbf{g}_{m}^{H}\mathbf{w}_{m}s(n)\!+\!
    \sqrt{p}\sqrt{\alpha}  q\mathbf{f}_{m}^{H}\mathbf{w}_{m}s(n)c(n) \big] \!+\!z(n),
  \end{equation}
where $ \alpha $ denotes the power reflection coefficient, $z(n)\sim\mathcal{CN}(0,\sigma^{2})$ is the additive 
white Gaussian noise. 
Based on the received signal $r(n)$ in (1), the receiver wishes to decode both the 
primary and secondary signals. To that end, since the backscatter 
link is typically much weaker than the direct link, the receiver may first decode 
the primary symbols $s(n)$, by treating the backscattered signals  as 
noise, whose power is
$\mathbb{E}[ p \alpha  |\sum_{m=1}^{M} q \mathbf{f}_{m}^{H}\mathbf{w}_{m} s(n)c(n)|^{2}]
=p \alpha  |q|^{2}|\sum_{m=1}^{M}   \mathbf{f}_{m}^{H}\mathbf{w}_{m}|^{2}$.
Therefore the signal-to-interference-plus-noise ratio (SINR) for 
decoding the primary information is  
\begin{equation}
\setlength\abovedisplayskip{2pt}
\setlength\belowdisplayskip{2pt}
  \gamma_{s}=\frac{p |\sum_{m=1}^{M} \mathbf{g}_{m}^{H}\mathbf{w}_{m}|^{2}}{ p\alpha |q|^{2}| \sum_{m=1}^{M} \mathbf{f}_{m}^{H}\mathbf{w}_{m}|^{2}+\sigma^{2}}.
\end{equation}

Note that due to the product of $c(n)$ and $s(n)$ in the second term of (1), 
the resulting noise for decoding $s(n)$ is no longer Gaussian. However, 
by using the fact that for any given noise power, Gaussian noise results 
in the maximum entropy and hence constitutes the worst-case noise 
\cite{HassibiB2003a, NeeserFD1993a},  
the achievable  rate of the primary signal in (1) is  
\begin{equation}
\setlength\abovedisplayskip{2pt}
\setlength\belowdisplayskip{2pt}
  R_{s}=\log_{2}(1+\gamma_{s}).
\end{equation}

After decoding the primary information, the first term in (1) can be subtracted 
from the received
signal before decoding the secondary symbols $c(n)$. The resulting signal is
\begin{equation}
\setlength\abovedisplayskip{2pt}
\setlength\belowdisplayskip{2pt}
  \hat{r}_{c}(n)=\sqrt{p}\sqrt{\alpha} q \sum\nolimits_{m=1}^{M} 
  \mathbf{f}_{m}^{H}\mathbf{w}_{m} s(n)c(n)  +z(n).
\end{equation}

Note that since $s(n)$ varies across different secondary symbols $c(n)$, 
(4) can be interpreted as  a fast-fading
channel,  whose instantaneous 
channel gain depends on $|s(n)|^{2}$ \cite{ZhangQ2019a}. With $s(n)\sim \mathcal{CN}(0,1)$,
its squared envelope follows an exponential distribution. 
Therefore, the ergodic rate of the backscatter communication (4) can 
be expressed as \cite{RLong2020a, TseD2005a}
\begin{equation}
\setlength\abovedisplayskip{2pt}
\setlength\belowdisplayskip{2pt}
  \begin{aligned}
    R_{c}  &=\mathbb{E}_{s} \Big[ \log_{2}\big(1+\frac{p\alpha|q|^{2}| \sum_{m=1}^{M} 
    \mathbf{f}_{m}^{H}\mathbf{w}_{m}|^{2}|s(n)|^{2} }{\sigma^{2}}\big)\Big]
    \\ &=\int_{0}^{\infty}\log_{2}(1+\beta x)e^{-x}dx
    \\ & =-e^{\frac{1}{\beta}}\mathrm{Ei}(-\frac{1}{\beta})\log_{2}e,
\end{aligned}
\end{equation}
where $\mathrm{Ei}(x)\triangleq \int_{-\infty}^{x}\frac{1}{u}e^{u}du$
is the exponential integral, and 
$\beta=\frac{p\alpha|q|^{2}| \sum_{m=1}^{M} \mathbf{f}_{m}^{H}\mathbf{w}_{m}|^{2}}{\sigma^{2}}$ 
is the average received SNR of the backscatter link. 

Note that the above analysis is based on the assumption of perfect CSI on 
$\mathbf{f}_{m}$, $\mathbf{g}_{m}$, and $q$. In practical wireless 
communication systems, these channels need to be acquired via e.g., pilot-based channel estimation. 
In the following, we propose the channel estimation methods for cell-free symbiotic 
radio systems and analyze the achievable rates
taking into account the channel estimation errors.

\section{Channel Estimation and Achievable Rate Analysis}

Similar to the extensively studied massive MIMO systems, efficient channel estimation 
for cell-free massive MIMO can be achieved by exploiting the uplink-downlink channel 
reciprocity \cite{KaltenbergerF2010a, MarzettaTL2010a}, i.e., the downlink channels 
can be efficiently estimated via uplink training. 
However, different from the existing cell-free massive MIMO systems \cite{Ngo2017a}, the channel 
estimation for cell-free symbiotic radio requires estimating not only the direct-link 
channels  $\mathbf {g}_{m}$, but also the backscatter channels $q$ and 
$\mathbf {f}_{m}, m=1,...,M$. To this end, in the following, we propose a two-phase 
based channel estimation method for cell-free symbiotic radio systems.  
In the first phase, 
pilot  symbols are sent by the receiver while muting the BD, so as to estimate
the direct-link channels $\mathbf{g}_{m}, m=1,...,M$. In the second phase, pilots are 
sent both by the receiver and the BD so that, together with the estimation of the direct-link 
channels, the cascaded backscatter channels $q\mathbf {f}_{m}$, are 
estimated.
\vspace{-0.2cm} 
\subsection{Direct-Link Channel Estimation  }

First, we discuss the uplink training-based estimation of the direct-link channels between 
the receiver and the $M$ APs. Denote by $\tau_{1}$ the length of the uplink training sequence, and
let $p_{t}$ be the training power. Further denote by $\boldsymbol{\varphi}_{1}
\in \mathbb{C}^{\tau_{1}\times 1}$ the pilot sequence, where $\|\boldsymbol{\varphi}_{1}\|^{2}= \tau_{1}$. 
The received training signals by the $N$ antennas of the $m$th AP over the $\tau_{1}$ symbol 
durations, which is denoted as $\mathbf{Y}'_{m}\in  \mathbb{C}^{ N \times \tau_{1}} $, can be written as
\begin{equation}
\setlength\abovedisplayskip{2pt}
\setlength\belowdisplayskip{2pt}
  \mathbf{Y}'_{m}=\sqrt{p_{t}}  \mathbf{g}_{m} \boldsymbol{\varphi}_{1}^{H} 
  +\mathbf{Z}'_{m},
  \qquad m=1,...,M,
\end{equation}
where $\mathbf{Z}'_{m}$ denotes the i.i.d CSCG noise with zero-mean and power $\sigma^2$. 
With the pilot sequence $\boldsymbol{\varphi}_{1}$ known at the APs,  
$\mathbf{Y}'_{m}$ can be projected to $\boldsymbol{\varphi}_{1}$, which gives  
\begin{equation}
\setlength\abovedisplayskip{2pt}
\setlength\belowdisplayskip{2pt}
  \check{\mathbf{y}}'_{m}=\frac{1}{\sqrt{p_{t}}} \mathbf{Y}'_{m} \boldsymbol{\varphi}_{1}= \tau_{1} \mathbf{g}_{m}
   +\frac{1}{\sqrt{p_{t}}} \hat{{\mathbf{z}}}'_{m},  
\end{equation}
where $\hat{\mathbf{z}}'_{m}=\mathbf {Z}'_{m} \boldsymbol{\varphi}_{1}$ is the resulting noise vector.  It can be shown that $\hat{\mathbf{z} }_{1,m}$
is i.i.d. CSCG noise with power $\tau_{1}\sigma^2$, i.e., $\hat {\mathbf {z}}'_{m} \sim 
\mathcal{CN}(\mathbf{0} , \tau_{1}\sigma^{2} \mathbf I_N)$.

With $\mathbf g_{m}$ being a zero-mean random vector, its linear minimum mean square 
error estimation (LMMSE),
denoted by $\hat{\mathbf{g}}_{m}\in  \mathbb{C}^{N \times 1}$,  is \cite{KayS1993a}
\begin{equation}
\setlength\abovedisplayskip{2pt}
\setlength\belowdisplayskip{2pt}
  \begin{aligned}
    \hat{\mathbf{g}}_{m}  &=\mathbb{E}[ \mathbf{g}_{m} \check{\mathbf{y}}^{'H}_{m} ] 
    \big(\mathbb{E}[\check{\mathbf{y}}'_{m} \check{\mathbf{y}}^{'H}_{m}]\big)^{-1}  \check{\mathbf{y}}'_{m} 
    \\ &=    \mathbf{R}_{\mathbf{g},m} (  \tau_{1}   \mathbf{R}_{\mathbf{g},m}+
    \frac{\sigma^2}{p_{t} }\mathbf{I}_{N})^{-1}\check{\mathbf{y}}_{m},
  \end{aligned} 
\end{equation} 
where $\mathbf{R}_{\mathbf{g},m}=\mathbb{E}[\mathbf{g}_{m}\mathbf{g}_{m}^{H}]$
denotes the   covariance matrix of $\mathbf g_{m}$. 
By further  decomposing the direct-link channel as $\mathbf g_{m}=\sqrt{b_{m}}\mathbf d_{m}$, 
with $b_{m}$ denoting the large-scale channel coefficient, and $\mathbf d_{m}\in \mathbb{C}^{N\times 1}$ denoting 
the zero-mean CSCG small-scale fading component, i.e., 
$ \mathbf{d}_m \sim \mathcal{CN}(\mathbf{0}, \mathbf I_N)$. Then $\mathbf g_{m}$ is 
CSCG distributed with covariance matrix 
$\mathbf{R}_{\mathbf{g},m}=b_{m} \mathbf I_{N}$, and thus LMMSE estimation is
also the optimal MMSE estimation. In this case, (8) can be simplified as
\begin{equation}
\setlength\abovedisplayskip{2pt}
\setlength\belowdisplayskip{2pt}
  \hat{\mathbf{g}}_{m}=\frac{p_{t}b_{m}}{p_{t} \tau_{1} b_{m}+\sigma^2}\check{\mathbf{y}}'_{m}.  
\end{equation}

It can be shown that $\hat{\mathbf g}_{m}$ follows the distribution
\begin{equation}
\setlength\abovedisplayskip{2pt}
\setlength\belowdisplayskip{2pt}
  \hat{\mathbf{g}}_{m} \sim \mathcal{CN}(\mathbf{0},\frac{e_{1} b_{m}^{2}}{1+e_{1}  b_{m}}\mathbf{I}_{N}),
\end{equation}
where we have defined the transmit  training energy-to-noise ratio (ENR) as 
$  e_{1}\triangleq \frac{p_t\tau_1}{\sigma^2}$.

Let $\tilde{\mathbf{g}}_{m}$  denote the channel estimation error of the $m$th AP, i.e.,
$\tilde{\mathbf{g}}_{m}=\mathbf{g}_{m}-\hat{\mathbf{g}}_{m}$. 
With MMSE estimation, it is known that $\tilde{\mathbf{g}}_{m}$ is uncorrelated with 
$\hat{\mathbf{g}}_{m}$ \cite{KayS1993a}, which follows the distribution
\begin{equation}
\setlength\abovedisplayskip{2pt}
\setlength\belowdisplayskip{2pt}
  \tilde{\mathbf{g}}_{m} \sim \mathcal{CN}(\mathbf{0},\frac{ b_{m}}{1+e_{1} b_{m}}\mathbf{I}_{N}).
\end{equation}

It is observed from (11) that as the transmit training ENR $e_{1}$ increases, the variance of 
the channel estimation error reduces, as expected.
\vspace{-0.2cm}
\subsection{Backscatter Channel Estimation  }
\setlength\belowdisplayskip{8pt}
With the estimation $\hat{\mathbf g}_{m}$ for the direct-link channels, in  the second phase, 
pilot symbols are sent from both the receiver and the
BD to estimate the cascaded backscatter channels $q\mathbf{f}_{m}$, $m=1,...,M$. 
Let $\tau_{2}$ denote the length of the training sequence in the second phase and 
$\boldsymbol{\varphi}_{2} \in  \mathbb{C}^{\tau_{2}\times 1}$ be the pilot sequence
sent by the  receiver, with $\|\boldsymbol{\varphi}_{2}\|^2=\tau_{2}$. The
received training signal by the $m$th AP can be written as
\begin{equation}
\setlength\abovedisplayskip{2pt}
\setlength\belowdisplayskip{2pt}
  \begin{aligned}
    \mathbf{Y}^{*}_{m}& = \sqrt{p_{t} \alpha}q  \mathbf{f}_{m} \boldsymbol{\varphi}_{2}^{H}+
                \sqrt{p_{t} } (\mathbf{\hat{g}}_{m}+\mathbf{\tilde{g}}_{m}) \boldsymbol{\varphi}_{2}^{H} 
                +\mathbf{Z}''_{m},
  \end{aligned}
\end{equation}
where $\mathbf{Z}''_{m}$ denotes the i.i.d. CSCG noise with power $\sigma^2$. Note that without loss 
of generality, we assume that the pilot symbols backscattered by the BD are all $1$. 
After subtracting  the terms related to the estimation  $ \hat{\mathbf{g}}_{m}$ 
of the direct-link channels from (12), we have
\begin{equation}
\setlength\abovedisplayskip{2pt}
\setlength\belowdisplayskip{2pt}
  \begin{aligned}
    \mathbf{Y}''_{m}& = \sqrt{p_{t} \alpha}q  \mathbf{f}_{m} \boldsymbol{\varphi}_{2}^{H}+
                \sqrt{p_{t} } \tilde{\mathbf{g}}_{m} \boldsymbol{\varphi}_{2}^{H} 
                +\mathbf{Z}''_{m}.
  \end{aligned}
\end{equation}

With $\boldsymbol{\varphi}_{2}$ known at the APs,  the projection of $\mathbf{Y}''_{m}$ 
after scaling by $\frac{1}{\sqrt{p_t \alpha}}$,  is
\begin{equation}
\setlength\abovedisplayskip{2pt}
\setlength\belowdisplayskip{2pt}
  \check{\mathbf{y}}''_{m}=\frac{1}{\sqrt{p_{t} \alpha}}\mathbf{Y}''_{m} \boldsymbol{\varphi}_{2}
  =     \tau_{2}   \mathbf{h}_{m} + \frac{\tau_{2}}{\sqrt{ \alpha}} \tilde{\mathbf{g}}_{m}
    +\frac{1}{\sqrt{p_{t}\alpha} } \hat{{\mathbf{z}}}''_{m} ,
\end{equation}
where we have defined  the cascaded backscatter channel as $\mathbf{h}_{m}\triangleq  
q  \mathbf{f}_{m}  $, and $\hat{{\mathbf{z}}}''_{m}\triangleq\mathbf {Z}''_{m} \boldsymbol{\varphi}_{2}$. 
It can be shown that $\hat{\mathbf{z} }''_{m} \sim 
\mathcal{CN}(\mathbf{0} , \tau_{2}\sigma^2 \mathbf I_N)$. 

Let $\mathbf{R}_{\mathbf{h},m}=\mathbb{E}[ \mathbf{h}_{m} \mathbf{h}_{m}^{H}]
$ denote the covariance matrix of the cascaded backscatter channel $\mathbf {h}_{m}$. 
Then the LMMSE estimation of $\mathbf {h}_{m}$ based on (14) is 
\begin{equation}
\setlength\abovedisplayskip{2pt}
\setlength\belowdisplayskip{2pt}
  \begin{aligned}
    \hat{\mathbf{h}}_{m}  &=\mathbb{E}[     \mathbf{h}_{m}  \check{\mathbf{y}}^{''H} _{m} ] 
    \big(\mathbb{E}[\check{\mathbf{y}}''_{m} \check{\mathbf{y}}^{''H}_{m}]\big)^{-1}  \check{\mathbf{y}}''_{m} 
    \\ &=     \mathbf{R}_{\mathbf{h},m} 
    ( \tau_{2}  \mathbf{R}_{\mathbf{h},m} + \frac{\tau_{2}}{  \alpha } 
    \mathbf{R}_{\tilde{\mathbf{g}},{m}}+
    \frac{\sigma^{2}}{p_{t}\alpha} \mathbf{I}_{N})^{-1}\check{\mathbf{y}}''_{m},
  \end{aligned} 
\end{equation} 
where $\mathbf{R}_{\tilde{\mathbf{g}},{m}}=\mathbb{E} [\mathbf{\tilde{g}}_{m}
\mathbf {\tilde{g}}_{m}^{H}]$ is the covariance matrix of $\tilde {\mathbf {g}}_{m}$.     

If the channel coefficients in $\mathbf {f}_{m}$ are i.i.d.  distributed with 
variance $\zeta_{m}$, we then have $\mathbf R_{\mathbf{h},m}=\mathbb{E}[|q|^2\mathbf f_{m} 
\mathbf f_m^H]=\upsilon_{m} \zeta_{m}\mathbf I_N=\epsilon_{m} \mathbf I_N$,
where $\upsilon_{m}=\mathbb{E}[|q|^2]$ and $\epsilon_{m}=\upsilon_{m} \zeta_{m}$.

As a result, (15) can be simplified as
\begin{equation}
\setlength\abovedisplayskip{2pt}
\setlength\belowdisplayskip{2pt}
  \hat{\mathbf{h}}_{m} =  \frac{\alpha p_{t} \epsilon_{m}}{\alpha p_{t} \tau_{2} \epsilon_{m} +
    \frac{p_{t}\tau_{2}b_{m}}{1+e_{1}b_{m}}+\sigma^2} \check{\mathbf{y}}''_{m}.
\end{equation}

Define the transmit training ENR in the second phase as $e_{2}\triangleq \frac{p_{t}\tau_{2}}{\sigma^2}$.
It then follows from (11) and (16) that   
\begin{equation}
\setlength\abovedisplayskip{2pt}
\setlength\belowdisplayskip{2pt}
  \begin{aligned}
    \mathbf{R}_{\hat{\mathbf{h}},m} &=\mathbb{E}[ \hat{\mathbf{h}}_{m}  \hat{\mathbf{h}}_{m}^{H}]
     =\frac{\alpha e_{2}   \epsilon_{m}^{2}} { \alpha e_{2}  \epsilon_{m}
    +\frac{e_{2}b_{m}}{1+e_{1}b_{m}}+1}\mathbf{I}_{N}.
  \end{aligned}
\end{equation}

Let $\tilde{\mathbf h}_{m}=\mathbf{h}_{m}-\hat {\mathbf h}_{m}$ 
denote the estimation error. We  have
\begin{equation}
\setlength\abovedisplayskip{2pt}
\setlength\belowdisplayskip{2pt}
  \begin{aligned}
    \mathbf{R}_{\tilde{\mathbf{h}},m} &
    \triangleq \mathbb{E}\big[(\mathbf {h}_{m}-\hat {\mathbf h}_{m})(\mathbf {h}_{m}-\hat {\mathbf h}_{m})^{H}\big]
      \\ &=\mathbf{R}_{\mathbf{h},m} - \mathbf{R}_{\hat{\mathbf{h}},m}
  \\ &=\frac{\epsilon_{m} (\frac{e_{2}b_{m}}{1+e_{1}b_{m}}+1)}
  {\alpha e_{2} \epsilon_{m}
  +\frac{e_{2}b_{m}}{1+e_{1}b_{m}}+1}\mathbf{I}_{N}.
  \end{aligned} 
\end{equation}

It follows from (11) that if $e_{1} \rightarrow \infty$, in which case the direct-link channel 
$\mathbf g_m$  is perfectly estimated without any error, the variance of the 
estimation error in (18) reduces to the same form as that in (11). 
\vspace{-2ex} 
\subsection{Achievable Rate Analysis}
\vspace{-1ex} 
In this subsection, we derive the  achievable  primary and secondary rates  
based on the  channel estimation $ \hat{\mathbf{g}}_{m} $ 
and $ \hat{\mathbf{h}}_{m}, m=1,...,M$, by taking into account the channel estimation errors. 
By substituting $\mathbf g_{m}=\hat {\mathbf g}_{m} + \tilde{\mathbf g}_{m}$ and 
$q\mathbf f_{m} =\hat {\mathbf h}_{m}+\tilde{\mathbf h}_{m}$ into (1), 
the received signal for information transmission can be written as
\begin{equation}
\setlength\abovedisplayskip{2pt}
\setlength\belowdisplayskip{2pt}
  \begin{aligned}
  r(n) &=    \sqrt{p} \sum\nolimits_{m=1}^{M} \big[(\hat{\mathbf{g}}_{m}+\tilde{\mathbf{g}}_{m})^{H}  {\mathbf{w}_{m}} s(n)+
  \\ & \sqrt{\alpha}(\hat{\mathbf{h}}_{m}+\tilde{\mathbf{h}}_{m})^{H} {\mathbf{w}_{m}} s(n)c(n) \big] +z(n). 
\end{aligned}
\end{equation}

For decoding  the primary signals $s(n)$, besides the interference from the 
backscatter symbols $c(n)$, the term caused by the channel estimation error 
$\tilde{\mathbf{g}}_{m}$ is also treated as noise \cite{YZeng2014a, HassibiB2003a}.  
Therefore, (19) can be  decomposed as
\begin{equation}
\setlength\abovedisplayskip{2pt}
\setlength\belowdisplayskip{2pt}
  r_{s}(n) = {\rm DS}'\cdot s(n)+ {\rm ER}+{\rm ST}+z(n),
\end{equation}
where ${\rm DS}', {\rm ER}$, and ${\rm ST}$ denote the desired signal, 
estimation errors and the secondary transmission signal respectively, which are given by
\begin{equation}
\setlength\abovedisplayskip{2pt}
\setlength\belowdisplayskip{2pt}
  {\rm DS}'=\sqrt{p} \sum\nolimits_{m=1}^{M}  \hat{\mathbf{g}}_{m} ^{H}  {\mathbf{w}_{m}} ,
\end{equation}
\begin{equation}
\setlength\abovedisplayskip{2pt}
\setlength\belowdisplayskip{2pt}
  {\rm ER}=\sqrt{p} \sum\nolimits_{m=1}^{M} \big( \tilde{\mathbf{g}}_{m}  
  +\sqrt{\alpha}\tilde{\mathbf{h}}_{m}c(n)\big)^{H} {\mathbf{w}_{m}}s(n) ,
\end{equation}
\begin{equation}
\setlength\abovedisplayskip{2pt}
\setlength\belowdisplayskip{2pt}
  {\rm ST}=\sqrt{p} \sum\nolimits_{m=1}^{M}  
  \sqrt{\alpha}\hat{\mathbf{h}}_{m} ^{H} {\mathbf{w}_{m}}s(n)c(n) .
\end{equation}

Therefore, the resulting SINR can be expressed as (24) shown at the top
of the next page, and the achievable rate is $  R_s=\log_2(1+\gamma_s)$.
\newcounter{TempEqCnt} 
\setcounter{TempEqCnt}{\value{equation}} 
\setcounter{equation}{23} 

\begin{figure*}[ht] 
  \begin{equation}
\setlength\abovedisplayskip{2pt}
\setlength\belowdisplayskip{2pt}
    \begin{aligned}
    \gamma_{s} & = \frac{|{\rm DS}'|^{2}}{\mathbb {E}_{s,c} \big[|{\rm ER} |^{2}\big]
    +\mathbb {E}_{s,c} \big[|{\rm ST} |^{2}\big]+\sigma^{2}}
    \\ &= \frac{| \sum_{m=1}^{M}\hat{\mathbf{g}}_{m}^{H}  {\mathbf{w}_{m}}|^{2}}
    { \mathbb{E}_{s,c}\big[|s(n)|^{2}| \sum_{m=1}^{M} \big( \tilde{\mathbf{g}}_{m}  
    \!+\!\sqrt{\alpha}  \tilde{\mathbf{h}}_{m}c(n)\big)^{H} {\mathbf{w}_{m}} |^{2}\big]  
    \! +\!\alpha\mathbb{E}_{s,c}\big[
     | \sum_{m=1}^{M} \!\hat{\mathbf{h}}_{m} ^{H} \!{\mathbf{w}_{m}}  |^{2} 
     |s(n)|^{2}|c(n)|^{2}\big] 
     \!+\! \frac{\sigma^{2}}{ p}  }.
     \\ &=\frac{| \sum_{m=1}^{M}\hat{\mathbf{g}}_{m}^{H}  {\mathbf{w}_{m}}|^{2}}
    {   \sum_{m=1}^{M}\sum_{l=1}^{M} {\mathbf{w}_{m}^{H}} ( \tilde{\mathbf{g}}_{m} 
    \tilde{\mathbf{g}}_{l}^{H}\!+\!
     \alpha  \tilde{\mathbf{h}}_{m}\tilde{\mathbf{h}}_{l}^{H}){\mathbf{w}_{l}}  
    \! +\!\alpha
     | \sum_{m=1}^{M} \!\hat{\mathbf{h}}_{m} ^{H} \!{\mathbf{w}_{m}}  |^{2}  
     \!+\! \frac{\sigma^{2}}{ p}  }.
  \end{aligned}
\end{equation}
\hrulefill  
\vspace{-0cm}
\end{figure*}  

Note that for any given channel estimations $\hat{\mathbf{g}}_{m} $ and 
$\hat{\mathbf{h}}_{m}$, since the channel estimation errors 
$\tilde{\mathbf{g}}_{m} $ and $\tilde{\mathbf{h}}_{m}$ are random, 
the SINR in (24) and hence its rate $R_s$ is random. 
By taking the expected achievable rate with respect to the random 
estimation errors $\tilde{\mathbf{g}}_{m} $ 
and $\tilde{\mathbf{h}}_{m}$, we have the result (25) shown at the top of the next page,
where $E=\sum_{m=1}^{M}\!\big[\frac{ b_{m}}{ 1+e_{1} b_{m} }\!
+\!\frac{\alpha \epsilon_{m} (\frac{e_{2}b_{m}}{1+e_{1}b_{m}}+1)}
{\alpha e_{2} \epsilon_{m}
+\frac{e_{2}b_{m}}{1+e_{1}b_{m}}+1}\big]\!+\!\frac{\sigma^{2}}{ p } \! $
accounts for the average channel estimation error and noise.  
Note that the inequality in (25) follows from   Jensen's inequality, and the fact that
 $\log_2(1+C/x)$ 
is a convex function for $x>0$.

\setcounter{TempEqCnt}{\value{equation}} 
\setcounter{equation}{24} 
\begin{figure*}[ht] 
  \begin{equation}
\setlength\abovedisplayskip{2pt}
\setlength\belowdisplayskip{2pt}
    \begin{aligned}
    \mathbb{E}[R_{s}]&=\mathbb{E}_{\tilde{\mathbf{g}}_{m}  
    , \tilde{\mathbf{h}}_{m}}\big[\log_{2}(1+\gamma_{s})\big] 
    \geq \log_{2}\Big(1+\frac{| \sum_{m=1}^{M}\hat{\mathbf{g}}_{m}^{H}  {\mathbf{w}_{m}}|^{2}}
    {  \mathbb{E}_{\tilde{\mathbf{g}}_{m}  
    , \tilde{\mathbf{h}}_{m}}\big[ \sum_{m=1}^{M}\sum_{l=1}^{M} {\mathbf{w}_{m}^{H}} ( \tilde{\mathbf{g}}_{m} 
    \tilde{\mathbf{g}}_{l}^{H}\!+\!
     \alpha  \tilde{\mathbf{h}}_{m}\tilde{\mathbf{h}}_{l}^{H}){\mathbf{w}_{l}}\big]  
    \! +\!\alpha
     | \sum_{m=1}^{M} \!\hat{\mathbf{h}}_{m} ^{H} \!{\mathbf{w}_{m}}  |^{2}  
     \!+\! \frac{\sigma^{2}}{ p}  }\Big)
     \\ &=\log_{2}\Big(1+\frac{| \sum_{m=1}^{M}\hat{\mathbf{g}}_{m}^{H}  {\mathbf{w}_{m}}|^{2}}
     { \sum_{m=1}^{M} {\mathbf{w}_{m}^{H}} (\mathbf{R}_{\tilde{\mathbf{g}},m}  + 
      \alpha\mathbf{R}_{\tilde{\mathbf{h}},m}  ) {\mathbf{w}_{m}} + \frac{\sigma^{2}}{ p} 
      +\alpha
      |\! \sum_{m=1}^{M}\!\hat{\mathbf{h}}_{m}^{H}  {\mathbf{w}_{m}} |^{2}  }\Big)
     \\ &=\log_{2}\Big(1+\frac{| \sum_{m=1}^{M}\hat{\mathbf{g}}_{m}^{H}  {\mathbf{w}_{m}}|^{2}}
     {E+\! \alpha
   |\! \sum_{m=1}^{M}\!\hat{\mathbf{h}}_{m}^{H}  {\mathbf{w}_{m}} |^{2} }\Big) 
  \end{aligned}
  \end{equation}
  \hrulefill  
  \vspace{-0.6cm}
\end{figure*} 
Next, we derive the achievable rate of the secondary signals $c(n)$. After  
decoding $s(n)$, the primary signals $s(n)$ can be subtracted from (19) based on  the 
estimated  channel $\hat{\mathbf{g}}_{m}$. The resulting signal is
\begin{equation}
\setlength\abovedisplayskip{2pt}
\setlength\belowdisplayskip{2pt}
  \begin{aligned}
   {r}_{c}(n)&= \sqrt{p} \sum\nolimits_{m=1}^{M} \big[\sqrt{\alpha}  ( \hat{\mathbf{h}}_{m}+\tilde{\mathbf{h}}_{m})^{H}  {\mathbf{w}_{m}}s(n)c(n) 
  \\ &+   \tilde{\mathbf{g}}_{m}^{H}  {\mathbf{w}_{m}} s(n)\big]+z(n).
  \end{aligned}
\end{equation}

By treating the terms caused by the channel estimation error 
$\tilde{\mathbf{g}}_{m}$ and $\tilde{\mathbf{h}}_{m}$ as noise, 
(26) can be decomposed as
\begin{equation}
\setlength\abovedisplayskip{2pt}
\setlength\belowdisplayskip{2pt}
  r_{c}(n)= {\rm DS}''\cdot c(n)+{\rm ER}+z(n),
\end{equation}
where  $ {\rm ER}$  denotes the
estimation errors given in (22), and ${\rm DS}''$ denotes the desired signal,  
which is given by
\begin{equation}
\setlength\abovedisplayskip{2pt}
\setlength\belowdisplayskip{2pt}
  {\rm DS}''=\sqrt{p} \sum\nolimits_{m=1}^{M}  \sqrt{\alpha}    \hat{\mathbf{h}}_{m}
  ^{H}  {\mathbf{w}_{m}}s(n) ,
\end{equation}
 
The resulting   SINR is  
\begin{equation}
\setlength\abovedisplayskip{2pt}
\setlength\belowdisplayskip{2pt}
  \begin{aligned}
    \gamma_{c} &\!=\!\frac{|{\rm DS}''|^{2}}{\mathbb{E}_{s,c}\big[|{\rm ER} |^{2}\big] +\sigma^{2}}
    \\ &\!=\!\frac{\alpha | \sum_{m=1}^{M}\hat{\mathbf{h}}_{m}^{H}  {\mathbf{w}_{m}}|^{2} |s(n)|^{2}}
    {\sum_{m=1}^{M}\sum_{l=1}^{M} {\mathbf{w}_{m}^{H}} ( \tilde{\mathbf{g}}_{m} 
    \tilde{\mathbf{g}}_{l}^{H}\!+\!
     \alpha  \tilde{\mathbf{h}}_{m}\tilde{\mathbf{h}}_{l}^{H}){\mathbf{w}_{l}}  \! 
     +  \!  \frac{\sigma^{2}}{ p}  },
  \end{aligned}
\end{equation}
and the achievable rate is $  R_c=\log_2(1+\gamma_c)$.

Note that different from (24), as the desired channel ${\rm DS}''$ also depends on the 
primary symbols $s(n)$, the SNR in (29) is a random variable that depends on 
both $|s(n)|^2$ and the channel estimation errors. Consider the expectation of $R_{c}$,
with the expectation taken with respect to both 
$|s(n)|^2$ and the channel estimation errors, we have
\begin{equation}
\setlength\abovedisplayskip{2pt}
\setlength\belowdisplayskip{2pt}
  \begin{aligned}
    &  \mathbb{E}[R_{c}] =\mathbb{E}_{\tilde{\mathbf{g}}_{m}  
    , \tilde{\mathbf{h}}_{m}, s}\big[\log_{2}(1+\gamma_{c})\big] 
    \\ &\geq \!\mathbb{E}_{s}\Big[\log_{2}\big(1\!+\!\frac{\alpha | \sum_{m=1}^{M}\hat{\mathbf{h}}_{m}^{H}  
    {\mathbf{w}_{m}}|^{2}|s(n)|^{2} }
    { \mathbb{E}_{\tilde{\mathbf{g}}_{m}  
    , \tilde{\mathbf{h}}_{m} }\big[\sum\limits_{m=1}^{M}\sum\limits_{l=1}^{M} {\mathbf{w}_{m}^{H}} ( \tilde{\mathbf{g}}_{m} 
    \tilde{\mathbf{g}}_{l}^{H}\!+\!
     \alpha  \tilde{\mathbf{h}}_{m}\tilde{\mathbf{h}}_{l}^{H}){\mathbf{w}_{l}} \big]\! 
     +  \!  \frac{\sigma^{2}}{ p}  }\big)\Big]
   \\ &=\mathbb{E}_{s}\Big[\log_{2}\big(1+\frac{ \alpha | \sum_{m=1}^{M}
   \hat{\mathbf{h}}_{m}^{H} {\mathbf{w}_{m}}|^{2} |s(n)|^{2} }
   {E}\big)\Big]
   \\ &=\int_{0}^{\infty}\log_{2}(1+\beta_{c} x)e^{-x}dx
    \\ &=-e^{\frac{1}{\beta_{c}}}\mathrm{Ei}(-\frac{1}{\beta_{c}})\log_{2}e,
  \end{aligned}
\end{equation}
where the inequality is obtained by applying Jensen's inequality to the 
convex function $\mathrm{log}_2(1+C/x)$, and
 $\beta_{c}=\frac{ \alpha | \sum_{m=1}^{M}
\hat{\mathbf{h}}_{m}^{H} {\mathbf{w}_{m}}|^{2}   }{E}$ represents the average
SNR for the secondary signals taking into account the channel estimation errors.

\vspace{-0.2cm}
\subsection{Weighted-MRT Beamforming}
\setlength\belowdisplayskip{8pt}
It can be observed from (25) and (30) that for cell-free symbiotic systems, 
the achievable primary and secondary communication rates depend on the 
transmit beamforming vectors $\mathbf{w}_m$. In particular, in order to maximize 
the primary communication link, all the $M$ APs set their beamforming 
vector $\{\mathbf w_m\}_{m=1}^M $ as the MRT beamforming vector matched to 
the estimated direct-link channel $\hat{\mathbf{g}}_m$,  which is
\begin{equation}
\setlength\abovedisplayskip{2pt}
\setlength\belowdisplayskip{2pt}
  \mathbf{ w}_{m}^{s}=\frac{\hat{\mathbf{g}}_{m}}
  {\Vert \hat{\mathbf{g}}_{m} \Vert}, \qquad m=1,...,M.
\end{equation}

On the other hand, to maximize the  secondary communication rate in (30),
$\mathbf{w}_m$ is set as the MRT beamformer matched to the estimated backscatter 
link $\hat{\mathbf{h}}_m$, which is
\begin{equation}
\setlength\abovedisplayskip{2pt}
\setlength\belowdisplayskip{2pt}
  \mathbf{ w}_{m}^{c}=\frac{\hat{\mathbf{h}}_{m}}
  {\Vert \hat{\mathbf{h}}_{m} \Vert}, \qquad m=1,...,M.
\end{equation}

In order to achieve a flexible trade-off between the primary and secondary 
communication rate, in this paper, we propose a low-complexity 
{\it weighted-MRT} beamforming scheme, where the transmit beamforming vector 
for each AP is set as
\begin{equation}
\setlength\abovedisplayskip{2pt}
\setlength\belowdisplayskip{2pt}
  \mathbf{w}_{m}=\kappa\big[\rho \mathbf{ w}^{s}_{m}+
  (1-\rho)\mathbf{ w}^{c}_{m}\big], \qquad m=1,...,M,
\end{equation}
where $0\leq \rho \leq 1$ is a weighting coefficient that controls 
the trade-off between the primary and secondary communication rate, 
and $\kappa$ is a power normalization factor to ensure $\|\mathbf{w}_{m}\|^2=1$ 
for any given $\rho$. By varying $\rho$ between 0 and 1, the achievable rate 
regions of the primary and secondary transmission   can be obtained. Note that 
\emph{weighted-MRT} beamforming is especially appealing 
for cell-free symbiotic radio systems, due to its low-complexity and scalability, since 
each AP can perform the beamforming locally with its own channel estimations 
$\hat{\mathbf{g}}_{m}$ and $\hat{\mathbf{h}}_{m}$, without having to exchange
the estimated CSIs among APs.

\section{Simulation Results}
In this section,  simulation results are provided to evaluate the performance of 
cell-free symbiotic radio systems. 
We set up a Cartesian coordinate system, where the BD is located at the 
origin (0,0), and the receiver is located at (5m, 0). 
Furthermore, we assume that  $M=16$ APs, each with $N=4$ antennas, are 
evenly spaced in a square area of size $750\mathrm{m} \times 750\mathrm{m} $, 
i.e., their locations correspond to the $4 \times 4$ grid points, with 
the x- and y-coordinates chosen from the set \{-375m, -125m, 125m, 375m\}.  
The channels of 
all communication links are independent, where the small-scale fading coefficients
follow the i.i.d. CSCG distribution with zero mean and unit variance. 
Furthermore, the large-scale channel gains of all links are  modeled as 
$b=\beta_{0}  d^{-\gamma}$,
where $\beta_{0}=(\frac{\lambda}{4\pi})^{2}$ is the reference channel gain 
with $\lambda=0.0857$m denoting the wavelength, $d$ represents the corresponding link 
distance, and $\gamma$ denotes the path loss 
exponent. We set $\gamma=2.7$ for the AP-to-BD and AP-to-receiver channels, and 
$\gamma=2.1$ for  BD-to-receiver channels. 
The power reflection coefficient is $\alpha=1$, and the 
transmitter-side SNR for both information and pilot transmission is set as 
$\frac{p}{\sigma^{2}}=\frac{p_{t}}{\sigma^{2}}=130$ dB,
which may correspond to $p=p_{t}=20$ dBm and $\sigma^2=-110$ dBm. 
The simulation results are obtained by taking the average values
over 1000 channel realizations.

Fig. 2 shows the achievable rate regions of the primary and secondary  rates 
with different uplink training lengths $\tau_{1}$, 
and hence different training ENR ${e_1=\frac{p_t\tau_1}{\sigma^2}}$,
while the pilot length in the second training phase  is fixed to 
$\tau_{2}=100$. Note that each point
of the curve corresponds to a primary-secondary rate pair with the 
weighted-MRT beamforming (33), by varying the weight $\rho$ from 0 to 1 with 
step size 0.1. It is observed from
Fig. 2 that with the training SNR $\frac{p_{t}}{\sigma^2}$ and training length 
$\tau_2$ fixed, the achievable rate regions critically depend on the 
training length $\tau_{1}$.
For  $\tau_1=1$, which corresponds to low training ENR $e_1$ in the first phase, 
the secondary communication rate is almost zero, regardless of the beamforming 
weight $\rho$. This can be explained by the fact that when  $e_1$ 
is low, there exists severe channel estimation error for the direct-link 
channel estimation, whose detrimental effect will be exacerbated  
for the estimation of the weaker cascaded backscatter channels in the second training phase. 
This thus severely limits the achievable rate of the secondary backscattering communication. 
As  $\tau_1$ increases to 100 so that both the direct-link and backscatter 
channels are  estimated more accurately, the rate region enlarges significantly.

By fixing $\tau_1=100$, Fig. 3 plots the achievable rate regions with 
different training lengths $\tau_{2}$ in the second phase. 
Similar to Fig. 2, it is observed from Fig. 3 that the rate region enlarges significantly 
as $\tau_{2}$ increases,
as expected. It is also interesting to note that with larger $\tau_2$, 
the minimum primary communication rate (corresponding to $\rho\!=\!0$) actually 
reduces. This can be explained by the fact that as the cascaded backscatter 
channels are estimated more accurately with larger $\tau_2$, it results in stronger 
interference  to the primary communication with MRT beamforming matched to 
the secondary link, which thus decreases the minimum primary rate. 
However, the maximum primary  rate ($\rho=1$) is almost unaffected 
by $\tau_2$.  By comparing Fig. 2 and Fig. 3, it is also observed that larger rate 
regions are achieved for $\tau_2\!=\!10$ and $\tau_2\!=\!1$ in Fig. 3 than its 
counterpart in Fig. 2. This implies that if the total training length 
$\tau_1\!+\!\tau_2$ is fixed, higher priority should be given to the  first  training  phase. This is expected since the estimation of the direct-link 
channels in the first phase  impacts  not only the primary communication rate, 
but also the  quality of the channel estimation of the backscatter channels. 
\begin{figure}[!t]
  \centering
  \setlength{\abovecaptionskip}{-0.1cm}
  \setlength{\belowcaptionskip}{-0.2cm}
\centerline{\includegraphics[height=2.3in, width=2.92in]{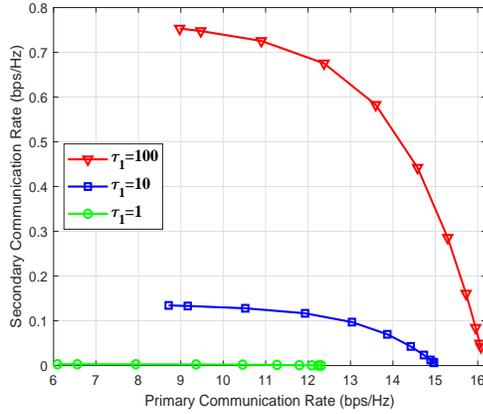}}
\caption{Achievable rate regions with different training lengths $\tau_1$ in 
the first phase, where  $\tau_{2}=100$.  }
\label{system model}
  \vspace{-0.4cm}
\end{figure}
\begin{figure}[!t]
	\centering
  \setlength{\abovecaptionskip}{-0.1cm}
  \setlength{\belowcaptionskip}{-0.2cm}
\includegraphics[height=2.3 in, width=2.92in]{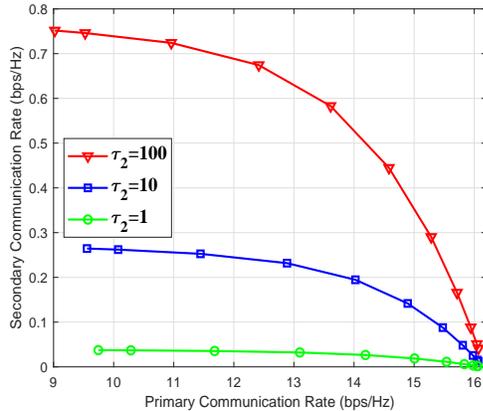}
\caption{Achievable rate regions with different training lengths $\tau_2$ in the 
second phase, where   $\tau_{1}=100$.  }
  \vspace{-0.6cm}
\end{figure}



\section{Conclusion}
In this paper, a novel cell-free symbiotic radio system was studied, 
in which a number of distributed APs cooperatively send primary information, 
while concurrently supporting the secondary backscattering communication. 
A two-phase uplink-training based channel
estimation method was proposed to estimate the direct-link channel and cascaded 
backscatter channel. 
Furthermore, a low-complexity weighted-MRT beamforming scheme 
was proposed to achieve a flexible trade-off between the primary and secondary
communication rate. 
Simulation results were provided to demonstrate the performance of 
the cell-free symbiotic
radio systems.

 

%


\vspace{0cm}
\section*{Acknowledgment}
\vspace{-1ex} 
This work was supported by the National Key R\&D Program
of China with Grant number 2019YFB1803400.

\ifCLASSOPTIONcaptionsoff
  \newpage
\fi

\end{document}